\documentclass[onecolumn,preprintnumbers,superscriptaddress, bibnotes,nofootinbib]{revtex4}

\usepackage{amsthm}
\usepackage{amsmath}
\usepackage{amssymb}
\usepackage{graphicx}
\usepackage{dsfont}
\usepackage{float}
\usepackage[dvipsnames]{xcolor}
\usepackage{mathrsfs}
\usepackage[colorlinks=true,citecolor=blue,urlcolor=black]{hyperref}
\usepackage{float}
\usepackage{afterpage}
\usepackage{adjustbox}
\usepackage{times}
\usepackage{tabularx}
\usepackage{multirow}

\begin{document}
	
	\title {Experiment on scalable multi-user twin-field quantum key distribution network}
	
	\author{Xiaoqing Zhong}
	\email{xzhong@physics.utoronto.ca}
	\affiliation{Center for Quantum Information and Quantum Control, Dept. of Physics, University of Toronto, Toronto, Ontario, M5S 1A7, Canada}
	
	\author{Wenyuan Wang}
	\altaffiliation[Current address: ]{Department of Physics, University of Hong Kong, Pokfulam Road, Hong Kong}
	\affiliation{Center for Quantum Information and Quantum Control, Dept. of Physics, University of Toronto, Toronto, Ontario, M5S 1A7, Canada}
	
	\author{Reem Mandil}
	\affiliation{Center for Quantum Information and Quantum Control, Dept. of Physics, University of Toronto, Toronto, Ontario, M5S 1A7, Canada}
	
	\author{Hoi-Kwong Lo}
	\affiliation{Center for Quantum Information and Quantum Control, Dept. of Physics, University of Toronto, Toronto, Ontario, M5S 1A7, Canada}
	\affiliation{Center for Quantum Information and Quantum Control, Dept. of Electrical \& Computer Engineering, University of Toronto, Toronto, Ontario, M5S 3G4, Canada}	
	\affiliation{Department of Physics, University of Hong Kong, Pokfulam Road, Hong Kong}	
	
	\author{Li Qian}
	\affiliation{Center for Quantum Information and Quantum Control, Dept. of Electrical \& Computer Engineering, University of Toronto, Toronto, Ontario, M5S 3G4, Canada}

	\begin{abstract}
	Twin-field quantum key distribution (TFQKD) systems have shown great promise for implementing practical long-distance secure quantum communication due to its measurement-device-independent nature and its ability to offer fundamentally superior rate-loss scaling than point-to-point QKD systems.  A surge of research and development effort in the last two years has produced many variants of protocols and experimental demonstrations. In terms of hardware topology, TFQKD systems interfering quantum signals from two remotely phase-locked laser sources are in essence giant Mach-Zehnder interferometers (MZIs) requiring active phase stabilization. Such configurations are inherently unsuitable for a TFQKD network, where more than one user-pair share the common quantum measurement station, because it is practically extremely difficult, if not impossible, to stabilize MZIs of largely disparate path lengths, a situation that is inevitable in a multi-user-pair TFQKD network. On the other hand, Sagnac interferometer based TFQKD systems exploiting the inherent phase stability of the Sagnac ring can implement asymmetric TFQKD, and are therefore eminently suitable for implementing a TFQKD network. In this work, we experimentally demonstrate a proof-of-principle multi-user-pair Sagnac TFQKD network where three user pairs sharing the same measurement station can perform pair-wise TFQKD through time multiplexing, with channel losses up to 58 dB, and channel loss asymmetry up to 15 dB. In some cases, the secure key rates still beat the rate-loss bounds for point-to-point repeaterless QKD systems, even in this network configuration. It is to our knowledge the first multi-user-pair TFQKD network demonstration, an important step in advancing quantum communication network technologies. 
	\end{abstract}
	\maketitle
	
	\subsection{Introduction}
		Quantum key distribution (QKD), an important quantum technology that allows the remote users to share encryption keys with information-theoretic security, has been well developed over the past few decades~\cite{R1,R2}. Various types of QKD protocols have been proposed and QKD experiments have been successfully performed over different systems~\cite{qkd_review}. A range of commercial QKD systems have also been developed and used in practical testbeds. More recently, a new type of QKD, called twin field quantum key distribution (TFQKD), has shown great promise for implementing practical long-distance secure quantum communication~\cite{tf-qkd_original}. For the conventional point-to-point QKD systems, the maximum key rate scales linearly with the channel transmittance $\eta$~\cite{plob,R21} without using quantum repeaters~\cite{QR}. For simplicity, we call this rate-lose limit the repeaterless bound. However, for TFQKD, the key rate scales linearly with $\sqrt{\eta}$, providing fundamentally superior rate-loss scaling than other repeaterless QKD systems. Moreover, TFQKD is inherently a measurement-device-independent QKD (MDIQKD)~\cite{mdiqkd}, where two remote users (conventionally called Alice and Bob) send encoded coherent states to an untrusted central node (conventionally called Charlie) who performs quantum measurements on the states. The measurements are only able to reveal the parity of the states sent by Alice and Bob, not the information encoded in the states, and the results are publicly announced. Hence, it is invulnerable to any attacks on detector side channels. Since the first proposal of TFQKD,  a surge of research and development effort in the last two years have produced many variants of protocols and security proofs~\cite{tfqkd1,tfqkd2,tfqkd3,tf-qkd_marcos,asy-tfqkd}. Meanwhile, a number of TFQKD experiments have demonstrated the feasibility of its application in long-distance quantum communication\cite{tf_exp1,tf_exp2,tf_exp3,tf_exp4,tf_exp5, tf_exp6,asy_tf_ex1}.
		
		Despite the rapid development, demonstrated TFQKD systems have only two participants exchanging keys (Alice and Bob), as is the case for all other conventional point-to-point QKD systems. To make TFQKD widely applicable in quantum communication in the future, one has to extend the two-user scenario to a multi-user case, that is, a network setting. QKD network~\cite{net} is an essential step towards building a global quantum internet which can enable more applications, such as cloud quantum computing~\cite{ibm}. There have been multiple QKD networks~\cite{net1,net2,net3,net4,net5,net6,net7,net8} built and tested all over the world, from small scales with a few users to large scales with more than a hundred users, such as the SECOQC network~\cite{net3}, the Tokyo QKD network~\cite{net6} and China's space-to-ground network~\cite{net8}. In these networks, different QKD protocols and technologies have been used. In particular, in China's space-to-ground network~\cite{net8}, two ground-to-satellite free-space QKD links are also integrated. All of these network building efforts have paved the way towards a global quantum internet. However, most existing QKD networks~\cite{net3,net5,net6,net8} are based on trusted central relays, which are undesirable for security. Any successful attacks of the central relays would break down the security of the network. There have been QKD networks using optical switches~\cite{net1,net2,net4} or using untrusted relays~\cite{net7}, but their key rates are limited by the repeaterless bound. In contrast, TFQKD can be employed in a network to solve both the security issue and key rate limit, due to its measurement-device-independent nature and its ability to outperform the repeaterless bound. Therefore, a TFQKD network would have a remarkable advantage over existing QKD networks.
		
		However, there are technical challenges to implement a TFQKD network. Because of the random phase fluctuations of signals over long distances, phase stabilization is required for the interference of coherent states in TFQKD. In most demonstrated TFQKD systems~\cite{tf_exp1,tf_exp3,tf_exp4,tf_exp5,tf_exp6}, two remote phase-locked laser sources are used to send quantum signals to the cnetral node for interference measurement, which is in essence a giant Mach-Zehnder interferometer (MZI). Such a configuration, with the use of active phase stabilization or post phase selection, is suitable for demonstrating a two-user TFQKD system. However, it is inherently unsuitable for an TFQKD network, where more than two users are involved and share the common central node. First of all, it would be impractical to have all the remote laser sources phase-locked, especially when large number of users are connected to the network. More importantly, it is extremely difficult, if not impossible, to stabilize an unbalanced MZI of largely disparate path lengths. In other words, such configuration would require that all the users have the same distances to the central node, while in a realistic network, the geographic distances between different users and the central node could be very different. On the other hand, there have been TFQKD systems using Sagnac interferometers which have the inherent phase stability~\cite{tf_exp2,asy_tf_ex1}. In the Sagnac interferometer configuration, the users share the same laser source that is held by the central node, removing the phase-locking process. Moreover, due to its common path nature, the Sagnac interferometer based TFQKD system has high tolerance for channel asymmetry~\cite{asy_tf_ex1}, and is therefore eminently suitable for implementing a TFQKD network. 

\begin{figure}[b]
		\includegraphics[width=0.7\linewidth]{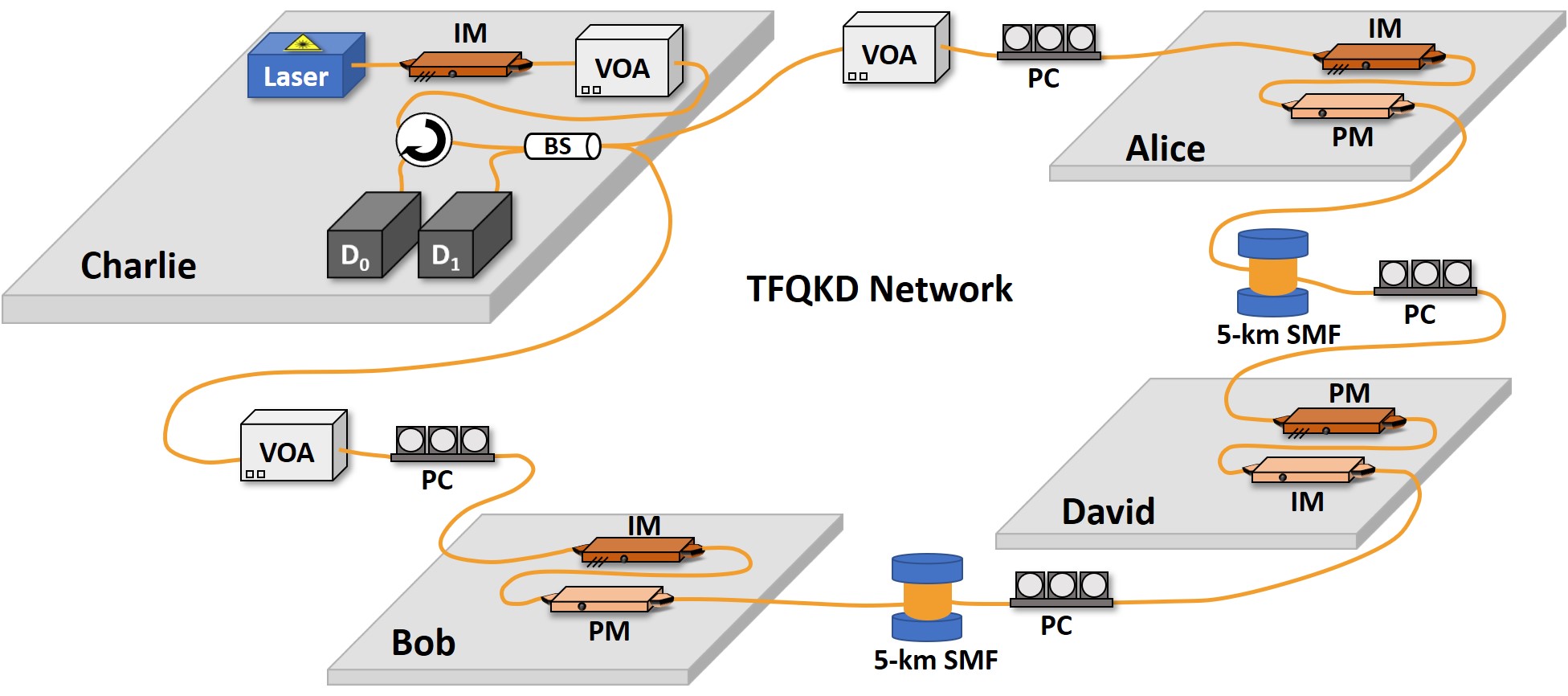}
		\caption{Schematic diagram of our experimental set-up of twin-field quantum key distribution network. On Charlie's station, that is located outside the Sganac loop, a continuous wave laser diode is used with the intensity modulator (IM) and variable optical attenuator (VOA) to generate the weak coherent pulses. Three users, Alice, Bob and David, are connected in the Sagnac loop. When any pair of users want to use the network to generate secret keys, Charlie launches the coherent pulses into the loop through a circulator (C) and a 50:50 beam splitter (BS). When the pulses arrive at the designated users, they use their intensity modulators (IM) and phase modulators (PM) to set the intensities and add the phases to the pulses. Note that there are clockwise and counterclockwise traveling pulses in the loop. Each active user only modulates one of the pulses, the one that has already traveled through the other active user, while letting the other pulse pass through without modulation. After the modulation, the pulses will be forwarded back to Charlie for measurement and will be detected by two single photon detectors $D_0$ and $D_1$. As a proof-of-principle demonstration, VOAs are inserted between Alice/Bob and Charlie to simulate the optical channel losses. Between Alice/Bob and David, there are 5 km single mode fibers (SMF). The polarization controllers (PC) are also used inside the loop for the polarization alignment.}
		\label{fig1}
	\end{figure}

		In this work, we, for the first time, experimentally demonstrate a proof-of-principle multi-user Sagnac TFQKD network.  As shown in FIG.~(\ref{fig1}), a Sagnac interferometer is applied in our TFQKD network. Three users, Alice, Bob and David, are connected in the loop.  One untrusted central node, Charlie, is located outside the loop and is equipped with a laser source and a pair of single photon detectors (SPDs). Any two of these users can perform pair-wise TFQKD at a given time slot. To run this network, Charlie sends weak coherent pulses into the fiber loop. The users encode their information into the pulses that are in their designated time slots and forward the modulated pulses back to Charlie for measurement. Compared with other QKD networks where the users have their own light sources and detectors, this topology of ours significantly reduces the cost for each user. It is also straightforward to add more users into the Sagnac TFQKD network. In our TFQKD network, channel losses between the central node and different users in our network could be either the symmetric or asymmetric. To optimize the key rates for different pairs of users, we choose different strategies in different situations. More specifically, a TFQKD protocol (called "CAL19") studied in Refs.~\cite{tf-qkd_marcos,tf_exp2} is applied when two users have the same channel loss to the central node, while an asymmetric version of CAL19 protocol studied in Refs.~\cite{asy-tfqkd,asy_tf_ex1} is adopted when two users have asymmetric loss to the central node. We show that our Sagnac interferometer based TFQKD network can allow users asymmetrically located in the ring network to share secret keys, even with channel loss asymmetry up to 15 dB. Without using quantum repeaters or trusted central relay, the secret key rate in some case can still beat the repeaterless bound.
	
	\subsection{Experiment}
		The experimental setup of our demonstration is shown in FIG.~(\ref{fig1}). On Charlie's station, a continuous wave laser diode is used with the intensity modulator and variable optical attenuator to generate the weak coherent pulses at a repetition rate of $10$ MHz with $900$ ps pulse width. Inside the Sagnac loop, there are three users, Alice, Bob and David. As a proof-of-principle demonstration, we use variable optical attenuators between Alice/Bob and Charlie to simulate the optical channel losses. Between Alice/Bob and David, there are 5 kilometers of single mode fiber. On each user's station, there are phase modulator and intensity modulator installed for information encoding. The polarization controllers are also used inside the loop for the polarization alignment. Here, our proof-of-principal demonstration has only the crucial components of information encoder for each user. For security concern, intensity monitors and filters should have been added on each user's station to exclude various side channel attacks by an eavesdropper. We would like to point out that these components can be easily included in our current setup. 
		
\begin{figure}
		\includegraphics[width=0.7\linewidth]{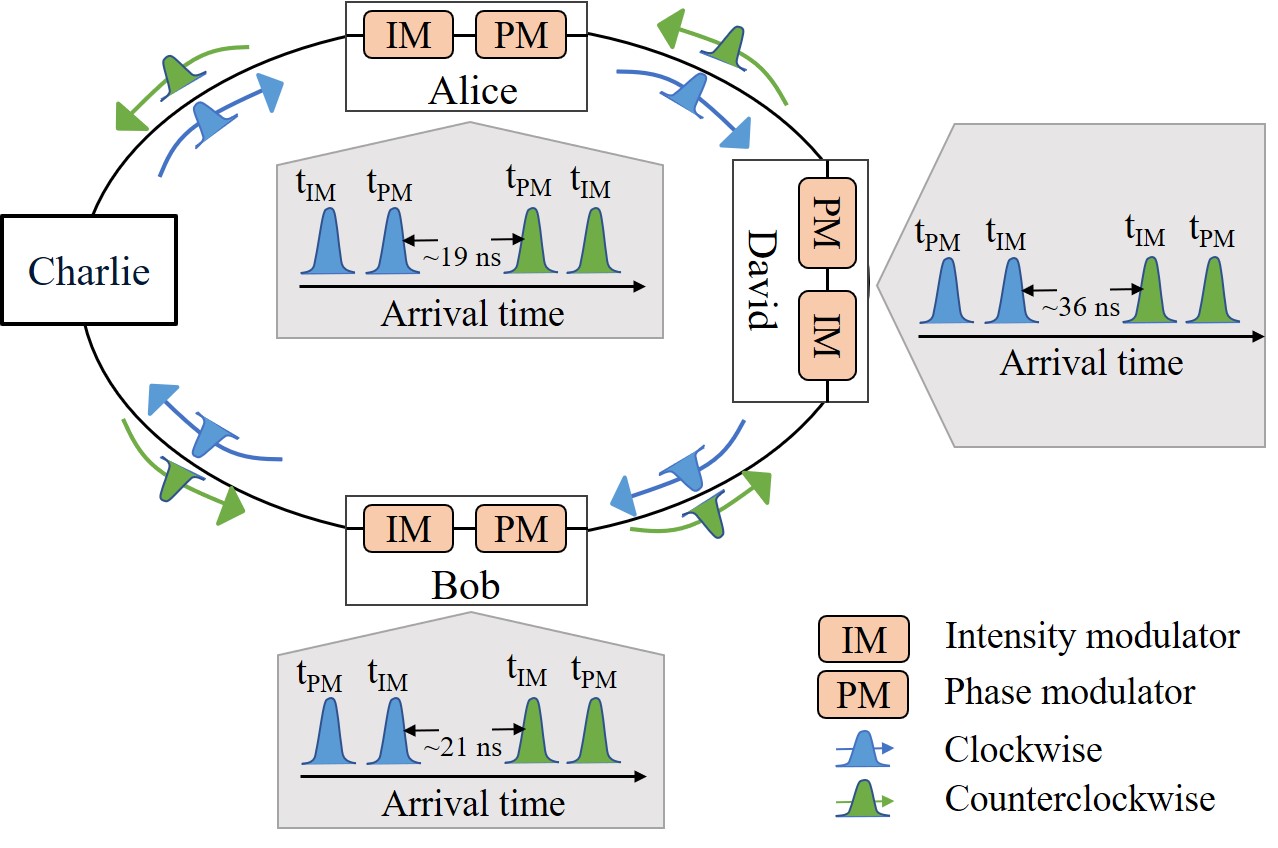}
		\caption{Illustration of different arrival times of clockwise and counterclockwise traveling pulses at different users' stations. Inside the loop, the pulse launched by Charlie at a rate of $10$ MHz are divided into two beams, one traveling in clockwise direction (marked in blue) while another one traveling in counterclockwise direction (marked in green). To avoid the collisions between clockwise and counterclockwise traveling pulses at any modulators, individual user can add a small amount of fiber at one end of his/her station so that the arrival times of two counter-propagating pulses at each user's station are different. On Alice's station, the counterclockwise traveling pulse arrives at Alice's phase modulator (PM) in about $19$ ns after the clockwise traveling pulse passing through Alice's PM. On Bob's station, the arrival time of the counterclockwise pulse at Bob's intensity modulator (IM) is about $21$ ns later than that of the clockwise pulse. On David's station, the difference between the arrival times of the clockwise and counterclockwise traveling pulses at David's IM is about $36$ ns. Note that it takes about $13$ ns for a pulse traveling from Alice's/Bob's IM to PM and about $15$ ns for a pulse traveling from David's IM to PM. The large difference of the arrival times guarantees that each user can only modulate the pulse traveling in one direction without making any changes to the other pulse, since the modulation window is only about $1$ ns. }
		\label{fig2}
	\end{figure}

		When any two users want to start the QKD protocol for key generation (the two users are called active users), Charlie launches weak coherent pulses into the loop through a circulator and a 50:50 beam splitter. As shown in FIG.~(\ref{fig2}), there are clockwise and counterclockwise traveling pulses in the loop. Each active user only modulates one of the pulses, the one that has already traveled through the other active user, while letting the other pulse pass through without modulation. For instance, when Alice and Bob are the active users, Alice (Bob) only modulates the pulses travelin g in counterclockwise (clockwise) direction. When the counter-propagating pulses are phase and intensity modulated by the active users and are forwarded back to Charlie, they interfere at the beam splitter and are measured by Charlie's two single photon detectors (ID220) $D_0$ and $D_1$ with a detection window of $900$ ps. The dark count probability of both detectors is about $7\times10^{-7}$. Note that when the protocol is running, the third inactive user will simply let the pulses go through his/her station without any modulation. We remark that this implementation does not compromise security, since the pulses will always undergo the fibers that are exposed to eavesdroppers. 
		 
		To mimic real network situations, we have both the symmetric and asymmetric channel losses (the loss between the user and the central node) for different pairs of users in our network. More specifically, Alice and Bob have the same channel losses, while the channel loss of David is 10 dB higher than the channel loss of Alice and 15 dB higher than the loss of Bob. When Alice and Bob use the network to share secret keys, they will run the CAL19 protocol~\cite{tf-qkd_marcos}. That is, Alice and Bob will randomly choose $x$ and $z$ bases for the pulses. For the pulses in $x$ basis, Alice and Bob randomly add a $0$ or $\pi$ phase and set the intensity to be the signal intensity $s$. For the pulses in $z$ basis, then Alice and Bob choose a random phase between $0$ and $2\pi$ and set the intensity to be one of the decoy intensity settings $\left\lbrace \mu,\nu,\omega\right\rbrace$. For the pair of users that have asymmetric channel losses (David\&Alice, David\&Bob), they will run the asymmetric CAL19 protocol~\cite{asy-tfqkd}. The difference between this implementation and the original CAL19 protocol is that, to compensate for the channel loss asymmetry to get a good interference visibility, the users will use asymmetric intensities for signal states in $x$ basis. As shown in Ref.~\cite{asy_tf_ex1}, with asymmetric intensities, the users can obtain a better key rate than of simply padding loss to make channels symmetric. For the two users' decoy states in $z$ basis, which are used for phase error rate estimation, the intensities can be either symmetric or asymmetric, since the channel asymmetry has little effects on the phase error rate~\cite{asy-tfqkd}.
		
		In our TFQKD network, with the use of Sagnac interferometer, the laser source and SPDs are required only by the central node Charlie. This configuration not only reduces the cost, but also remarkably simplifies the operations for the users in the network. Since   interference of coherent state is used in TFQKD, the coherent states sent out by the users should have matched frequency and matched global phase. As mentioned before, if the users have their own lasers, a process of phase locking of these individual lasers is necessary. In addition, the phases of the coherent states reaching Charlie need to be actively stabilized. In contrast, for Sagnac interferometry, neither phase locking nor phase stabilization is necessary, since a single laser is shared by the users and the interfering pulses travel through a common path\footnote{The phase stability of a Sagnac interferometer is dependent on the fiber length of the loop. As long as the phase fluctuations of signals over the light transit time through half the loop is small, the automatic phase stability is guaranteed. Ref.~\cite{tf_exp2} has estimated that for a loop length of $300$ km, the Sagnac-interferometer configuration without active phase stabilization is still applicable for TF-QKD.}. The only operation left for the users is the information encoding (intensity and phase modulations). Therefore, users can be also easily added into or removed from our network. 
				
		Even though there are pulses traveling bidirectionally in our setup, each user will only modulate the pulses in one direction. Therefore, it is important for us to ensure that the clockwise and counterclockwise traveling pulses would not collide at any modulators of the users inside the loop. To achieve this condition, fiber segments with well calibrated lengths are added on different users' stations to avoid the pulse collision at each user's station. As shown in FIG.~(\ref{fig2}), the difference of arrival times of the counter-propagating pulses at each user's station is at least $19$ ns in our experiment, which adequately ensures that the modulation applied to the pulse in one direction will not affect the pulse in the other direction. (Note that the modulation window is only $1$ ns.) It is also very important to guarantee that the active users will impose modulation only when the designated pulses arrive. In our demonstration, all the intensity and phase modulators are synchronized and driven by a high-speed multi-channel arbitrary waveform generator (Keysight M8195). The delay time of the electrical signal to each modulator is carefully adjusted such that the designated pulses will exactly fall into the modulation window when they pass through the modulators.
		
	\subsection{Results}
		In our demonstration, different pairs of users have exchanged secret keys through the network. By adjusting the variable optical attenuator, we have varied the channel losses for different users and tested the key rates over different cases. For Alice and Bob who have the same channel loss, one overall loss point is tested, that is $40.12$ dB. Note that the overall loss represents the total channel losses of two users. For David and Alice who always have $10.00$ dB channel loss asymmetry, the tested overall losses are $50.00$ dB and $58.00$ dB. For David and Bob who have $15.00$ dB channel loss asymmetry, they have exchanged keys over both $43.16$ dB and $51.16$ dB. All the signal intensities ($s$) and decoys states ($\mu, \nu, \omega$) used in our tests are listed in Tab.~(\ref{tab1}). For Alice and Bob, they perform the original CAL19 protocol~\cite{tf-qkd_marcos} to share secret keys. As shown in Tab.~(\ref{tab1}), Alice and Bob have the symmetric intensities for both their signal and decoy states. While for another two pairs of users, to compensate for the channel loss asymmetry, they will use different signal intensities as suggested in the asymmetric CAL19 protocol~\cite{asy-tfqkd}. The user with higher channel loss always sends out stronger signals than the other user. But note that the ratio of the optimal signal intensities of the two users slightly deviates from the inverse ratio of the their channel transmittance~\cite{asy-tfqkd}. For the decoy states, the users can either choose the same or asymmetric decoy intensities. In our demonstration, because of the loop configuration, we simply choose the ratio of decoy intensities of two users to be the inverse ratio of their channel transmittance. For maximizing the secure key rate, all the intensities used in our experiments are close to but not exactly the same as the optimal intensities. For all pairs of users in our network, they always send $1\times10^{11}$ pulses to the central node during each test. The probabilities of sending signal states or decoys states are also listed in Tab.~(\ref{tab1}).
		
\begin{figure}[b]
		\includegraphics[width=0.8\linewidth]{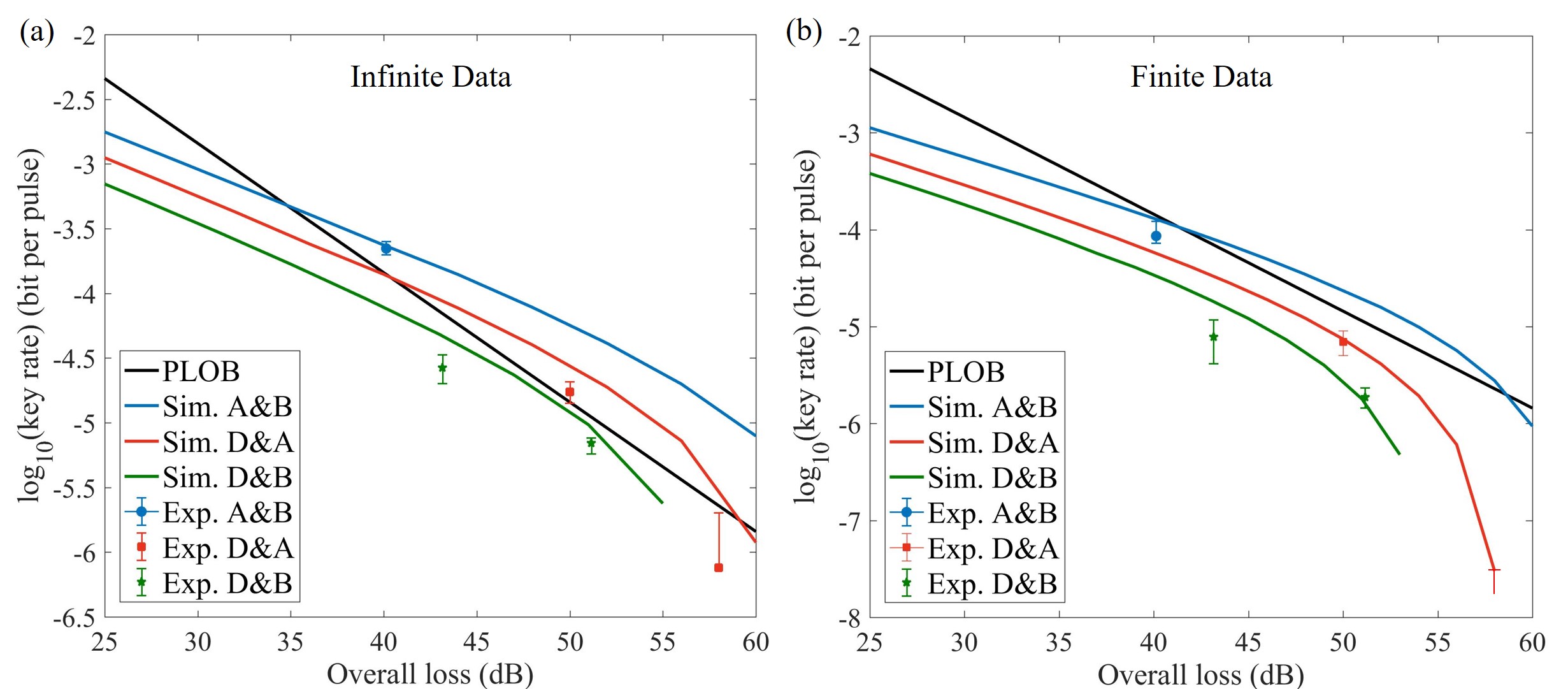}
		\caption{Log-log plot of the key rates of different users in the network as a function of the overall loss. The overall loss represents the sum of the channel losses of two users to the central node Charlie. Different pairs of users have tested the network and the experimental key rates are calculated in both a) infinite-data scenario and b) finite-data scenario.  For Alice and Bob who have the symmetric channel losses, one overall loss point is tested, that is $40.12$ dB. The experimental key rates are shown as blue circles. For David and Alice, who have $10.00$ dB Channel loss asymmetry, they have tested the network over both $50.00$ dB and $58.00$ dB overall channel losses. Their experimental key rates are represented by red squares. For David and Bob, who have $15.00$ dB Channel loss asymmetry, they have tested the network over both $43.16$ dB and $51.16$ dB overall channel losses. Their experimental key rates are shown as green stars. The vertical bar of each data point indicates the best and worst key rates when intensity fluctuation is taken into consideration. Note that at the overall loss is $58.00$ dB, the worst key rate for David and Alice is $0$ in the infinite-data scenario. While in the finite-data scenarios, both the average key rate and the worst key rate are $0$. The solid curves show the simulated key rates for different pairs of users. The black solid line is the PLOB bound~\cite{plob}, which is one representative of the repeaterless bound.}
		\label{fig3}
	\end{figure}

		The observed quantum bit error rate (QBER) for each test is shown in Tab.~(\ref{tab2}). As the overall loss and channel loss asymmetry between the users increase, the observed QBER also increases. We remark that in our demonstration, no active phase or polarization compensation is applied during each test. Due to the long-term stability of our TFQKD network system, the largest QBER observed in our experiment is still less than $5\%$. The secret key rate (bit per pulse) for each test is calculated based on the experimental gains and QBERS and is listed in Tab.~(\ref{tab2}) as well. In our key rate analysis, we consider both the infinite-data and finite-data scenarios. Meanwhile, we also take the intensity fluctuation into consideration and calculate the best and the worst key rates for each test. The results are also shown as scattered points in FIG.~(\ref{fig3}), which is a plot of the secret key rate in logarithmic scale as a function of the overall channel loss. To compare the performance of our TFQKD network with the rate-loss limit of conventional QKD networks without trusted central relays, we also plot out one representative of the repeaterless bound, that is the PLOB bound~\cite{plob} (represented by the solid black line). As shown in FIG.~(\ref{fig3}a) where infinite-data size is assumed\footnote{Note that here the simulation curves are based on intensities optimized against asymptotic key rate, while the experimental key rates are calculated with intensities optimized for finite-data case. That is to say, the experimental intensity settings are not optimal in the infinite-data case. Therefore, overall, the experimental key rates are a bit lower than the simulation curves in FIG.~(\ref{fig3}a).}, the secret key rate for Alice and Bob is $2.227\times10^{-4}$ at $43.16$ dB (equivalent to $216$ km), which is consistent with the simulation. More importantly, this key rate is significantly higher than the PLOB bound, even when the worse key rate is considered. For the other two pairs of users, due to the channel loss asymmetry, their secret key rates would be lower than the key rates of Alice and Bob. However, with our strategy of using asymmetric intensities, they can still obtain optimal key rates compared with other compensation strategies~\cite{asy_tf_ex1}. As shown in FIG.~(\ref{fig3}a), even with $15.00$ dB channel loss asymmetry, David and Bob can still successfully share secret keys over both overall losses. For the high loss of $51.16$ dB (equivalent to $256$ km), the key rate is as high as $6.946\times10^{-6}$, which is close to the PLOB bound. For David and Alice who have $10.00$ dB channel loss asymmetry, their secret keys rate at $50.00$ dB loss (equivalent to $250$ km) is $1.728\times10^{-5}$, still higher than the PLOB bound. When the overall loss is as high as $58.00$ dB (equivalent to $290$ km), the pulses that are launched into the loop by Charlie are the strongest compared with the previous cases. So, the backscattering noise is relatively high and affects the performance of our system. This is reflected in the large range of the secret key rate.  Even though, the best key rate David and Alice can obtain in our test is as high as $2.014\times10^{-6}$. Even when the finite-size effect is considered, as indicated in FIG.~(\ref{fig3}b), the experimental key rates are consistent with the simulations, showing that our TFQKD network still allows different users to share secret keys. As explained before, the only exception is that, when the overall loss between David and Alice is $58.00$ dB, due to the backscattering noise, the average key rate is $0$. But when intensity fluctuation is considered, secret keys can still be shared between David and Alice at a positive key rate.
		
	\subsection{Discussion}
		In summary, we have proposed and demonstrated the first TFQKD network with three users and one untrusted central node. A Sagnac loop configuration is used in our network, which not only reduces the cost for the users, but also makes our network adaptable for adding more users). In our proof-of-principle demonstration, the network has been tested by different pairs of users. Meanwhile, to reflect real world channel asymmetries, both symmetric and asymmetric channel losses between users have been tested in this demonstration. To achieve high secure key rates in both symmetric and asymmetric channel-loss scenarios, different versions of CAL19 TFQKD protocol have been used. The experimental results show that, the users in our network is able to share secret keys with a rate that is higher than the repeaterless bound without trusted relays. Moreover, for the users with different channel losses, without the help of any trusted central node, they can successfully exchange secret keys over an overall loss as high as $58$ dB which is equivalent to a fiber distance of $290$ km. We remark that while we have chosen a particular CAL19 protocol in our experimental demonstration, the basic principle of our design of a Sagnac interferometer based TFQKD network works well for other class of protocols such as the sending or not sending protocol~\cite{tfqkd2}.
		
		In this work, we focus on showing the feasibly of building a scalable TFQKD network that enables multi-pair of users to share encryption keys through an untrusted central node. Therefore, as a proof-of-principle demonstration, variable optical attenuators, rather than long fibers, are used in our setup. For future research, a field test of such a TFQKD network would be interesting. The variable optical attenuators should be replaced by long fibers. Moreover, extra components, such as intensity monitors, attenuators and filters, should be added on each user's station to against the attacks from any eavesdropper. For long distance applications, the main limitation of the design of our TFQKD network is the backscattering noise of the fibers. As discussed in Ref.~\cite{asy_tf_ex1}, there are viable strategies to alleviate the backscattering issue. One way to limit the backscattering noise is to lower the intensity of the pulses launched into the loop. Bidirectional amplifiers can be inserted between the users to compensate for the fiber loss. Another way to deal with the backscattering issue is to use bursts of pulses. One can design the timing and the duration of the bursts such that the signals arrive at the detectors within the period during which the backscattering noise decays.
		
		Another interesting question would be how to enable different pairs of users in the network to exchange secret keys simultaneously. In our current implementation, each pair of users are allowed to perform TFQKD to share keys through the network at a given time slot. In other words, time multiplexing is applied. This implementation is very simple but adequate for our purpose of demonstrating the feasibility of a TFQKD network. To improve the efficiency of the network, that is to enable different pairs of users to simultaneously generate keys, wavelength multiplexing technique could be applied. For instance, with the use of wavelength division multiplexing, different pairs of users can use different wavelength channels to transmit signals simultaneously. Note that the central node then would need multiple pairs of detectors for the signals in different wavelength channels, which will add complexity to the system. In conclusion, while further improvements could be done to our network, our demonstration has shown the practicability of a Sagnac interferometer based TFQKD network, which not only removes the security dependence on the trusted central relays of existing QKD networks with trusted nodes, but also beats the distance limitation of repeaterless QKD networks. We hope that with more research, TFQKD network can be well developed as a promising approach in advancing quantum communication network technologies.

	\subsection{Acknowledgment}
		This work is supported by funding from NSERC, MITACS, CFI, ORF, Royal Bank of Canada, Huawei Technology Canada, CRCEF, and the University of Hong Kong start-up grant.

	\begin{table*}
		\begin{tabular}{ccccccccccc}
			\hline\hline
			 & User & \parbox{1.2cm}{Channel\\ loss (dB)} & $s$ & $\mu$ & $\nu$ & $\omega$ & $P_s$ & $P_{\mu}$ & $P_\nu$ & $P_\omega$\\\hline 
			 
			 \multirow{2}{1cm}{Pair 1}& Alice & $20.06$ & {$0.0242\pm0.0002$} & {$0.1009\pm0.0003$} & $0.0241\pm0.0002$&{$(2.1\pm0.3)\times10^{-4}$} & \multirow{2}{*} {$0.86$} & \multirow{2}{*} {$0.035$} & \multirow{2}{*} {$0.033$} & \multirow{2}{*} {$0.072$} \\		
			 & Bob & $20.06$ & {$0.0242\pm0.0002$} &{$0.1009\pm0.0003$} &{$0.0241\pm0.0002$} &{$(2.1\pm0.3)\times10^{-4}$} & & & &\\	\hline			 

			 \multirow{4}{1cm}{Pair 2}& David & $30.00$ &{$0.0350\pm0.0014$} &{$0.798\pm0.002$} &{$0.168\pm0.002$} &{$(3.50\pm0.63)\times10^{-4}$}& \multirow{2}{*} {$0.84$} & \multirow{2}{*} {$0.032$} & \multirow{2}{*} {$0.036$} & \multirow{2}{*} {$0.092$} \\		
			& Alice & $20.00$ &{$0.00421\pm0.00003$} &{$0.0798\pm0.0002$} &{$0.0168\pm0.0002$} &{$(3.50\pm0.63)\times10^{-5}$} & & & &\\
						
			\cline{2-11}& David & $34.00$&{$0.0246\pm0.0007$} &{$0.656\pm0.002$} &{$0.1403\pm0.0008$} &{$(1.17\pm0.19)\times10^{-3}$} & \multirow{2}{*} {$0.32$} & \multirow{2}{*} {$0.125$} & \multirow{2}{*} {$0.147$} & \multirow{2}{*} {$0.408$} \\
			& Alice & $24.00$&{$0.00436\pm0.00004$} &{$0.0656\pm0.0002$} &{$0.01403\pm0.00008$} &{$(1.17\pm0.19)\times10^{-4}$} & & & &\\\hline

			 \multirow{4}{1cm}{Pair 3}& David & $29.08$ &{$0.050\pm0.002$} &{$0.769\pm0.002$} &{$0.1587\pm0.0009$} &{$(7.3\pm0.6)\times10^{-4}$}& \multirow{2}{*} {$0.83$} & \multirow{2}{*} {$0.034$} & \multirow{2}{*} {$0.043$} & \multirow{2}{*} {$0.093$} \\		
			& Bob & $14.08$ &{$0.00228\pm0.000001$} &{$0.02517\pm0.00006$} &{$0.00502\pm0.00003$} &{$(2.3\pm0.2)\times10^{-5}$} & & & &\\
			
			\cline{2-11}& David & $33.08$&{$0.0200\pm0.0002$} &{$0.783\pm0.002$} &{$0.1533\pm0.0008$} &{$(3.1\pm1.2)\times10^{-4}$} & \multirow{2}{*} {$0.71$} & \multirow{2}{*} {$0.050$} & \multirow{2}{*} {$0.065$} & \multirow{2}{*} {$0.175$} \\
			& Alice & $18.08$&{$0.00113\pm0.00002$} &{$0.02476\pm0.00006$} &{$0.00485\pm0.00003$} &{$(0.98\pm0.38)\times10^{-5}$} & & & &\\		 
			\hline\hline			
		\end{tabular}
		\caption{List of intensities and sending probabilities of signal and decoy states used in our experiments. $s$ is the signal intensity. $\mu$, $\nu$ and $\omega$ are the decoy intensities. For the pair of users who have the same channel loss, both their signal intensities and decoy intensities are symmetric. For the pair of users who have different Channel losses, their signal and decoy intensities are asymmetric. The probabilities of sending signal state ($P_s$) or one of the decoy states ($P_{\mu}, P_{\nu}, P_{\omega}$) are always the same for a pair of users. }
		\label{tab1}
	\end{table*}

\begin{table*}
	\begin{tabular}{@{\extracolsep{4pt}}ccccccccccc@{}}
		\hline\hline
		&\multirow{2}{*}{User} & \multirow{2}{*}{\parbox{1.2cm}{Overall\\ loss (dB)}} & \multicolumn{2}{c}{QBER}&\multicolumn{3}{c}{Key rate (infinite-data)}& \multicolumn{3}{c}{Key rate (finite-data)}\\
		\cline{4-5} \cline{6-8} \cline{9-11} \noalign{\smallskip}& & & $D_0$ & $D_1$ & Mean & Best & Worst & Mean & Best & Worst \\\hline 
		\multirow{1}{*}{Pair 1} & A-B & $40.12$ & {$0.26\%$} & {$0.25\%$} & $2.227\times10^{-4}$ & $2.502\times10^{-4}$ & $1.976\times10^{-4}$ &	$8.587\times10^{-5}$ & $1.003\times10^{-4}$	& $7.261\times10^{-5}$\\\hline	
		
		\multirow{2}{*}{Pair 2}& D-A & $50.00$ & {$1.75\%$} & {$1.73\%$} & $1.728\times10^{-5}$ & $2.063\times10^{-5}$ & $1.411\times10^{-5}$ &	$6.961\times10^{-6}$ & $9.012\times10^{-6}$	& $5.047\times10^{-6}$\\
		& D-A & $58.00$ & {$4.34\%$} & {$4.60\%$} & $7.551\times10^{-7}$ & $2.014\times10^{-6}$ & $0$ & $0$ & $3.262\times10^{-8}$	& $0$\\\hline
		
		\multirow{2}{*}{Pair 3}& D-B & $43.16$ & {$2.15\%$} & {$1.96\%$} & $2.653\times10^{-5}$ & $3.334\times10^{-5}$ & $2.010\times10^{-5}$ &	$7.827\times10^{-6}$ & $1.174\times10^{-5}$	& $4.144\times10^{-6}$\\
		& D-B & $51.16$ & {$3.80\%$} & {$3.73\%$} & $6.946\times10^{-6}$ & $7.644\times10^{-6}$ & $5.729\times10^{-6}$ &	$1.872\times10^{-6}$ & $2.321\times10^{-6}$	& $1.450\times10^{-6}$\\			 	 
		\hline\hline			
	\end{tabular}
	\caption{List of observed quantum bit error rate (QBER) and secret key rate for different pairs of users. Here, A represents Alice, B is for Bob and D is for David. The QBERs observed in both detector $D_0$ and detector $D_1$ are given. The secret key rates are calculated in both infinite-data scenario and finite-data scenario. For the latter case, the total data size is $1\times10^{11}$. In both scenarios, intensity fluctuations are considered to find the best and worst key rates. }
	\label{tab2}
\end{table*}


\begin{thebibliography}{}
		\bibitem{R1}Bennett, C. H., and Brassard, G., Quantum cryptography: Public key distribution and coin tossing. \textit{In Proc. of IEEE Int. Conf. on Comp., Syst. and Signal Proc.}, Bangalore, India (1984).
		\bibitem{R2}Ekert, A. K., Quantum cryptography based on Bell’s theorem. \textit{Physical Review Letters}, 67(6), 661 (1991).
		
		\bibitem{qkd_review} Xu, F., Ma, X., Zhang, Q., Lo, H.-K., and Pan, J. W.. Secure quantum key distribution with realistic devices. \text{Reviews of Modern Physics}, 92(2), 025002 (2020).
		
		\bibitem{tf-qkd_original}	Lucamarini, M., Yuan, Z. L., Dynes, J. F., and Shields, A. J., Overcoming the rate–distance limit of quantum key distribution without quantum repeaters. \textit{Nature}, 557(7705), 400 (2018).
		
		\bibitem{R21}	Takeoka, M., Guha, S., and Wilde, M. M., Fundamental rate-loss tradeoff for optical quantum key distribution. \textit{Nature Communications}, 5, 5235 (2014).
		\bibitem{plob}   Pirandola, S., Laurenza, R., Ottaviani, C., and Banchi, L., Fundamental limits of repeaterless quantum communications. \textit{Nature Communications}, 8, 15043 (2017).
						
		\bibitem{QR} Briegel, H. J., Dür, W., Cirac, J. I., and Zoller, P., Quantum repeaters: the role of imperfect local operations in quantum communication. \textit{Physical Review Letters}, 81(26), 5932 (1998).
		
		\bibitem{mdiqkd}Lo, H.-K., Curty, M., and Qi, B., Measurement-device-independent quantum key distribution. \textit{Physical Review Letters}, 108(13), 130503 (2012).
		
		\bibitem{tfqkd1} Ma, X., Zeng, P., and Zhou, H., Phase-matching quantum key distribution. \textit{Physical Review X}, 8(3), 031043 (2018).
		\bibitem{tfqkd2}	Wang, X. B., Yu, Z. W., and Hu, X. L., Twin-field quantum key distribution with large misalignment error. \textit{Physical Review A}, 98(6), 062323 (2018).
		\bibitem{tfqkd3}	Lin, J., and L\"utkenhaus, N., Simple security analysis of phase-matching measurement-device-independent quantum key distribution. \textit{Physical Review A}, 98(4), 042332 (2018).
		\bibitem{tf-qkd_marcos} Curty, M., Azuma, K., and Lo, H.-K., Simple security proof of twin-field type quantum key distribution protocol. \textit{npj Quantum Information}, 5(1), 1-6 (2019).
		\bibitem{asy-tfqkd} Wang, W. and Lo H.-K., Simple Method for Asymmetric Twin-Field Quantum Key Distribution. {\textit{New Journal of Physics}, 22, 013020 (2019)}.
		
		\bibitem{tf_exp1} Minder, M., Pittaluga, M., Roberts, G. L., Lucamarini, M., Dynes, J. F., Yuan, Z. L., and Shields, A. J., Experimental quantum key distribution beyond the repeaterless secret key capacity. \textit{Nature Photonics}, 1 (2019).			
		\bibitem{tf_exp2} Zhong, X., Hu, J., Curty, M., Qian, L., and Lo, H.-K., Proof-of-principle experimental demonstration of twin-field type quantum key distribution. \textit{Physical Review Letters}, 123(10), 100506 (2019).		
		\bibitem{tf_exp3} Liu, Y., Yu, Z. W., Zhang, W., Guan, J. Y., Chen, J. P., Zhang, C., \textit{et al.}, Experimental Twin-Field Quantum Key Distribution Through Sending-or-Not-Sending. \textit{Physical Review Letters}, 123(10), 100505 (2019).
		\bibitem{tf_exp4} Wang, S., He, D. Y., Yin, Z. Q., Lu, F. Y., Cui, C. H., Chen, W., Zhou, Z., Guo, G. C., and Han, Z. F., Beating the fundamental rate-distance limit in a proof-of-principle quantum key distribution system. \textit{Physical Review X}, 9(2), 021046 (2019).
		\bibitem{tf_exp5} Fang, X. T., Zeng, P., Liu, H., Zou, M., Wu, W., Tang, Y. L. \textit{et al.}, Implementation of quantum key distribution surpassing the linear rate-transmittance bound. \textit{Nature Photonics}, 1-4 (2020).
		\bibitem{tf_exp6} Chen, J. P., Zhang, C., Liu, Y., Jiang, C., Zhang, W., Hu, X. L. \textit{et al.}, Sending-or-Not-Sending with Independent Lasers: Secure Twin-Field Quantum Key Distribution over 509 km. \textit{Physical Review Letters}, 124(7), 070501 (2020).
		\bibitem{asy_tf_ex1} Zhong, X., Wang, W., Qian, L., and Lo, H.-K. Proof-of-principle experimental demonstration of twin-field quantum key distribution over optical channels with asymmetric losses. \textit{npj Quantum Information}, 7(1), 1-6 (2021).
		
		\bibitem{net}Townsend, P. D., Phoenix, S. J. D., Blow, K. J., and Barnett, S. M., Design of quantum cryptography systems for passive optical networks. \textit{Electronics Letters}, 30(22), 1875-1877 (1994).
		\bibitem{ibm}Castelvecchi, D., IBM's quantum cloud computer goes commercial. \textit{Nature News}, 543(7644), 159, (2017).
		\bibitem{net1}Elliott, C., Colvin, A., Pearson, D., Pikalo, O., Schlafer, J., and Yeh, H., Current status of the DARPA quantum network. \textit{Proc. SPIE 5815, Quantum Information and Computation III} (2005).
		\bibitem{net2}Chen, W., Han, Z. F., Zhang, T., Wen, H., Yin, Z. Q., Xu, F. X. \textit{et al.}, Field experiment on a “star type” metropolitan quantum key distribution network. \textit{IEEE Photonics Technology Letters}, 21(9), 575-577 (2009). 
		\bibitem{net3}Peev, M., Pacher, C., Alléaume, R., Barreiro, C., Bouda, J., Boxleitner, W., Debuisschert, T., Diamanti, E., Dianati, M., Dynes, J.F. and Fasel, S., The SECOQC quantum key distribution network in Vienna. \textit{New Journal of Physics}, 11(7), p.075001 (2009).	
		\bibitem{net4} Wang, S., Chen, W., Yin, Z.Q., Zhang, Y., Zhang, T., Li, H.W., Xu, F.X., Zhou, Z., Yang, Y., Huang, D.J. and Zhang, L.J., Field test of wavelength-saving quantum key distribution network. \textit{Optics Letters}, 35(14), 2454-2456 (2010).
		\bibitem{net5}Stucki, D., Legre, M., Buntschu, F., Clausen, B., Felber, N., Gisin, N., Henzen, L., Junod, P., Litzistorf, G., Monbaron, P. and Monat, L., Long-term performance of the SwissQuantum quantum key distribution network in a field environment. \textit{New Journal of Physics}, 13(12), 123001 (2011).
		\bibitem{net6}Sasaki, M., Fujiwara, M., Ishizuka, H., Klaus, W., Wakui, K., Takeoka, M., Miki, S., Yamashita, T., Wang, Z., Tanaka, A. and Yoshino, K.,  Field test of quantum key distribution in the Tokyo QKD Network. \textit{Optics Express}, 19(11), 10387-10409 (2011).
		\bibitem{net7}Tang, Y. L., Yin, H. L., Zhao, Q., Liu, H., Sun, X. X., Huang, M. Q. \textit{et al.},  Measurement-device-independent quantum key distribution over untrustful metropolitan network. \textit{Physical Review X}, 6(1), 011024 (2016).
		\bibitem{net8}Chen, Y. A., Zhang, Q., Chen, T. Y., Cai, W. Q., Liao, S. K., Zhang, J., Chen, K., Yin, J., Ren, J. G., Chen, Z. and Han, S. L., An integrated space-to-ground quantum communication network over 4,600 kilometres. \textit{Nature}, 589(7841), 214-219 (2021).			
			
	\end{thebibliography}
\end{document}